\RequirePackage[bookmarksnumbered,unicode]{hyperref}
\documentclass[sigconf]{aamas} 
\setcopyright{ifaamas}
\acmConference[AAMAS '23]{Proc.\@ of the 22nd International Conference on Autonomous Agents and Multiagent Systems (AAMAS 2023)}{May 29 -- June 2, 2023}{London, United Kingdom}{A.~Ricci, W.~Yeoh, N.~Agmon, B.~An (eds.)}
\copyrightyear{2023}
\acmYear{2023}
\acmDOI{}
\acmPrice{}
\acmISBN{}

\acmSubmissionID{38}
\usepackage{algorithm2e}

\usepackage[inline]{enumitem}

\usepackage{subcaption}
\usepackage{booktabs}
\usepackage{url}
\usepackage{flushend}
\usepackage{xcolor}
\definecolor{labelcolor}{cmyk}{0.22,0.10,0.10,0.10}
\definecolor{listbackgroundcolor}{cmyk}{0.10,0.10,0.05,0.05}
\definecolor{listbackgroundcolorlight}{rgb}{0.91,0.92,0.94}

\usepackage{listings,multicol}
\usepackage{tikz}
\usepackage{standalone}
\usetikzlibrary{calc}
\usetikzlibrary{fit}
\usetikzlibrary{positioning}

\newcommand{\ul}{\ulcorner}
\newcommand{\ur}{\urcorner}
\newcommand{\inn}{\ensuremath{\ul\mathsf{in}\ur}\xspace}
\newcommand{\out}{\ensuremath{\ul\mathsf{out}\ur}\xspace}
\newcommand{\nil}{\ensuremath{\ul\mathsf{nil}\ur}\xspace}

\newcommand{\true}{\textbf{T}}
\newcommand{\false}{\textbf{F}}

\usepackage[scr=boondoxo,scrscaled=1.05]{mathalfa}
\newcommand{\schema}[1]{#1}
\newcommand{\association}[1]{\ensuremath{\widehat{#1}}}
\newcommand{\partt}[1]{\ensuremath{\widehat{#1}}}
\newcommand{\sender}{\ensuremath{\mathscr{s}}}
\newcommand{\receiver}{\ensuremath{\mathscr{r}}}

\newcounter{ronecount}
\newcounter{rtwocount}
\newcounter{rthrcount}

\newcommand{\fsl}{\textsl}
\newcommand{\fsc}{\textsc}
\newcommand{\fsf}[1]{{\footnotesize{\textsf{#1}}}}

\newcommand{\fn}{\textsf}

\newcommand{\mname}[1]{\fsl{#1}}
\newcommand{\pname}[1]{\fsl{#1}}
\newcommand{\rname}[1]{\textsc{#1}}
\newcommand{\val}[1]{\texttt{#1}}
\newcommand{\paraname}[1]{\fsf{#1}}

\usepackage{fontawesome}

\usepackage{mathpartir}

\DeclareMathAlphabet{\mathcal}{OMS}{cmsy}{m}{n}

\DeclareRobustCommand{\nUmErAL}[1]{#1}

\hypersetup{pdfborder={\ 0\ 0\ 0}}

\usepackage{proof}
\newtheorem{theorem}{Theorem}

\newtheorem{definition}{Definition}
\newenvironment{proof-sketch}{\noindent\textit{Proof Sketch.}}{\hfill$\square$\vspace{1em}}

\usepackage{mdframed}
\begin{document}

\title{Kiko: Programming Agents to Enact Interaction Protocols}

\author{Samuel H.~Christie V}
\affiliation{%
  \institution{North Carolina State University}
  \city{Raleigh}
  \state{NC}
  \country{USA}
}
\email{schrist@ncsu.edu}
\orcid{0000-0003-1341-0087}

\author{Munindar P.~Singh}
\affiliation{
  \institution{North Carolina State University}
  \city{Raleigh}
  \state{NC}
  \country{USA}
}
\email{mpsingh@ncsu.edu}
\orcid{0000-0003-3599-3893}

\author{Amit K.~Chopra}
\affiliation{
  \institution{Lancaster University}
  \city{Lancaster}
  \country{UK}
}
\email{amit.chopra@lancaster.ac.uk}
\orcid{0000-0003-4629-7594}

\begin{abstract}
Realizing a multiagent system involves implementing member agents who interact based on a protocol while making decisions in a decentralized manner.  Current programming models for agents offer poor abstractions for decision making and fail to adequately bridge an agent's internal decision logic with its public decisions.  

We present \emph{Kiko}, a protocol-based programming model for agents. To implement an agent, a programmer writes one or more \emph{decision makers}, each of which chooses from among a set of valid decisions and makes mutually compatible decisions on what messages to send. 
By completely abstracting away the underlying communication service and by supporting practical decision-making patterns, Kiko enables agent developers to focus on business logic. We provide an operational semantics for Kiko and establish that Kiko agents are protocol compliant and able to realize any protocol enactment.
\end{abstract}



\pagestyle{fancy}
\fancyhead{}

\maketitle

\section{Introduction}
\label{sec:introduction}
Enterprise and other applications, e.g., in business and healthcare, involve interactions between social entities such as humans and organizations \cite{EIM92} based on technical resources such as databases.  
A sociotechnical system (STS) involves social and technical entities \cite{Pitt+12:axiomatization,TIST-13-Governance} and provides a useful abstraction for such applications.
Today, an STS is implemented using a conceptually central service through which its entities interact. 
In contrast, we address the challenges of implementing a decentralized multiagent system (MAS) to realize an STS.  Here, each principal maps to an agent; the agents interact with each other via \emph{asynchronous} messaging.  

The messages sent by an agent represent its public \emph{decisions}. For example, a \mname{Quote} by \rname{Seller} (for an item and price) represents a decision by it; an \mname{Accept} (of some \mname{Quote}) sent by \rname{Buyer} represents a decision of \rname{Buyer}; and so on.  To coordinate their decisions, the agents rely on an \emph{interaction protocol}. By specifying the constraints on messaging, a protocol specifies the constraints on decision making between the agents in a MAS.  For example, a \pname{Purchase} protocol in the above-introduced e-business setting may specify that the price is offered by the seller, and payment is required for delivery.

A protocol is specified abstractly with reference to roles to be adopted by agents in a multiagent system.  Implementing an agent according to a role means fleshing out the role with private (internal) decision logic that results in messages being emitted, that is, decisions being made \cite{TSE-05}.  For example, suppose agents \val{Bob} and \val{Sally} play \rname{Buyer} and \rname{Seller}, respectively, in \pname{Purchase}.  \val{Sally}'s decision logic may be to send  \mname{Quote}s with lower prices to repeat buyers. \val{Bob}'s decision logic may be to \mname{Accept} a \mname{Quote} if the price fits within its budget.  Such decision logic is the essence of an agent. 

Supporting the common desire \cite{little:microservices:2017,blog:serverless-business-logic:2021} for programming models that separate business logic from other components---and combating complexity in agent communication \cite{AC-Manifesto-13}, in general---proves challenging.  Traditional protocol languages \cite{AAMAS-Choice-09,Odell+2001,Honda+08, Winikoff+18:HAPN,Ferrando+19:enactability} specify message ordering, which limits flexibility \cite{JAIR-20:Langeval}. JADE \cite{Bergenti+20:JADE,Bellifemine+07:JADE}, a programming model for multiagent systems, is noteworthy for its early support for FIPA protocols \cite{FIPA-IP}; however, the FIPA approach is long outdated \cite{Computer-98} and the FIPA protocols are limited to a few patterns of interaction specified in terms of message ordering.  Agent-oriented programming models such as Jason \cite{Bordini+Huber-10:Jason} and JaCaMo \cite{Boissier+13:JaCaMo} provide cognitive abstractions for encoding an agent's internal reasoning but do not support protocols. Existing commitment-based approaches \cite{Gunay+15:generating, Winikoff-commitment-07} either rely on centralized commitment stores \cite{Baldoni+19:artifacts} or do not adequately address operationalizing asynchronous communication \cite{AAAI-Enact-08}; some approaches map the problem to protocols \cite{IJCAI-17:Tosca,AAAI-20:Clouseau}---and hence within the scope of this paper.
Traditional agent-oriented methodologies \cite{Cernuzzi+04:Gaia,Winikoff+Padgham-05,Cossentino+10:ASPECS} emphasize and incorporate protocols as design abstractions. However, the protocol specifications in these approaches are informal (usually UML interaction diagrams), which rules out protocol-based software abstractions for engineering agents. 
In a nutshell, today we lack a protocol-based programming model for agents that supports flexible, decentralized decision making via asynchronous messaging. 
 
Our contribution, \emph{Kiko}, addresses this gap. Specifically, Kiko advances a novel decision-oriented programming model that enables structuring and implementing agents based on the protocol roles they play.  Kiko's fundamental abstraction is that of a \emph{decision maker}, a construct for capturing the decision logic that selects and makes a set of decisions from those currently available.  The agent developer's primary task is to write the set of decision makers.  

Kiko guarantees an agent's compliance with the roles its plays. Kiko supports practical decision-making patterns that challenge other approaches, including correlation, cross-enactment reasoning, emission sets, and multiprotocol reasoning. Notably, in providing a decision-based programming interface, Kiko abstracts away the communication service that transports messages between agents.  In particular, decision making in Kiko avoids having to deal with the order in which messages are received. Actual message emission is also handled transparently in the programming model.

In addition, we contribute a formalization of the programming model and prove its soundness and completeness with respect to possible protocol enactments. We also present an optimized compliance-checking method and establish its validity.

\section{Information Protocols Introduced}
A protocol-based programming model for agents presumes a language in
which to specify protocols.
We adopt BSPL \cite{AAMAS-BSPL-11}, a declarative protocol language that
eschews the specification of message ordering and instead specifies
information constraints.

\begin{lstlisting}[label={BSPL:Purchase},caption={The \pname{Purchase} protocol.}]
Purchase {
  roles Buyer, Seller
  parameters out ID key, out item, out price, out done
  
  Buyer -> Seller: RFQ[out ID key, out item]
  Seller -> Buyer: Quote[in ID key, in item, out price]
  Buyer -> Seller: Buy[in ID key, in item, in price, out done]
  Buyer -> Seller: Reject[in ID key, in price, out done]
}
\end{lstlisting}

An information protocol in BSPL specifies the roles, messages between roles, and information constraints that define which message emissions are valid.
Information causality captures information dependencies: what information must or must not be known by an agent playing a role to be able to send a message.
Information integrity captures consistency in distributed settings: there cannot be two messages sent with conflicting information in the same protocol enactment.
Given the local store of an agent (its history of message observations), an agent can send any message that satisfies the specified causality and integrity constraints.

Listing~\ref{BSPL:Purchase} illustrates the main ideas of information protocols. 
It specifies a purchase protocol to be enacted by agents playing roles \rname{Buyer} and \rname{Seller}.
\pname{Purchase} composes message schemas, each with its sender and receiver roles and information parameters.
For example, \mname{RFQ} is from \rname{Buyer} to \rname{Seller} and its parameters are \paraname{ID} and \paraname{item}.
A concrete message instance associates the parameter names with value bindings, e.g., binding \paraname{ID} to a UUID and \paraname{item} to ``ball.''

To support information integrity, some parameters in a message schema are annotated \emph{key}, e.g., \paraname{ID} in all the messages of Listing~\ref{BSPL:Purchase}.
A tuple of bindings for the key parameters of a message schema uniquely identifies both an instance of the schema and the enactment to which it belongs, in which all nonkey parameters may have at most one binding.
For example, say \mname{RFQ} occurs with bindings [\paraname{ID}: \val{10}, \paraname{item}: \val{ball}].
Then, a \mname{Quote} with [\paraname{ID}: \val{10}, \paraname{item}: \val{hat}, \paraname{price}: 10] would violate integrity because for the same binding of \paraname{ID} there are different bindings of \paraname{item}.
Conversely, \mname{Quote} with [\paraname{ID}: \val{11}, \paraname{item}: \val{hat}, \paraname{price}: 10] satisfies integrity despite the different binding for \paraname{item} because it has a different binding of the key \paraname{ID}.

In a message schema, every message parameter is adorned \inn, \out, or \nil.
Adornments capture information causality constraints for the emission of an instance of a schema; \inn parameters must be known from prior communications (they are causal dependencies); \out parameters and \nil parameters must \emph{not} be known, but \out parameters are bound in the emission. For example, in Listing~\ref{BSPL:Purchase}, \rname{Seller} must know \paraname{item} before it can send \mname{Quote}, and in doing so produces a binding for \paraname{price}.

Knowledge of a parameter exists in the context of some binding for the associated key.
After receiving an \mname{RFQ} with bindings [\paraname{ID}: \val{10}, \paraname{item}: \val{ball}], \rname{Seller} knows that in the enactment \paraname{ID}=\val{10} \paraname{item} is bound to \val{ball}, and can produce a binding of \paraname{price} by sending \mname{Quote}.

Integrity and causality apply to protocols generally.  In \pname{Purchase} in Listing~\ref{BSPL:Purchase}, all protocol parameters are adorned \out in the protocol parameter line, meaning that each enactment of \pname{Purchase} as identified by the \paraname{ID} generates bindings for all of them.  Further, the parameter line enables composition with other protocols.

\section{The Kiko Programming Model}
\label{sec:pm}

We introduce the architectural basis for the programming model, followed by examples that illustrate its features.


\begin{figure}[htb!]
\centering
\resizebox{0.60\columnwidth}{!}{%
%
\begin{tikzpicture}[>=stealth]

\tikzstyle{emptybox}=[draw=none,fill=none,font=\small\sffamily,minimum height=15]

\tikzstyle{pbox}=[draw, rectangle, sharp corners,align=center,font=\sffamily,fill=gray!40,rectangle,anchor=center,minimum height=6ex,minimum width=20ex,inner sep=2]

\tikzstyle{abox}=[draw, rectangle, sharp corners,align=center,font=\sffamily,fill=gray!40,rectangle,anchor=center,minimum height=6ex,minimum width=20ex,inner sep=2]

\matrix (swiml) [draw=none,fill=none,rounded corners,row sep=30,
    column sep=5] {
      \node[pbox] (m) {MAS Info}; &
      \node[pbox] (d) {Decision Makers};  \\
     & \node[abox,draw=none] (a) {Protocol Adapter}; \\
     & \node[abox,draw=none,fill=gray!10] (c) {Communication\\ Service}; \\
};
\draw[->] (m) |- node[emptybox,above right=0.05] {Config} (a.west) ;

\draw[->] ($(d.south) + (-0.2,0)$) -- node[emptybox,left] {Attempts} ($(a.north) + (-0.2,0)$) ;
\draw[->]  ($(a.north) + (0.2,0)$) -- node[emptybox,right] {Forms} ($(d.south) + (0.2,0)$);
\draw[<->]  (a) -- node[emptybox, left=0.05] {Instances} (c);
\end{tikzpicture}

}
\caption{The Kiko agent architecture.}
\label{fig:kiko-agent}
\end{figure}
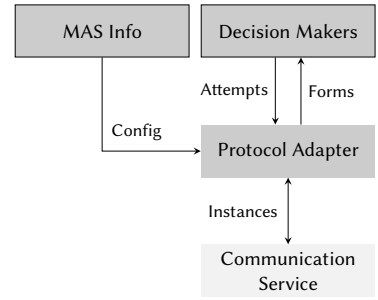

Figure~\ref{fig:kiko-agent} shows the main components of the agent architecture as focused on enacting protocols.
The MAS Info and Decision Makers are components provided by the agent programmer (indicated by the border).  The Protocol Adapter is a generic component provided by Kiko that understands information protocols and provides an API for plugging in Decision Makers. The adapters of all agents collectively achieve a coordination service and assimilate information received from messages \cite{Atal97}.
The Communication Service is anything that provides asynchronous messaging between agents. Our implementation uses UDP, which is unordered and unreliable (lossy).

An information protocol constrains only the \emph{emission} of messages by agents, based on its causal dependencies.
This means that ordered delivery, as provided by TCP or a message queue, is not required for correctly enacting a protocol.
Further, message reception is idempotent, so messages can be retransmitted to enact a protocol reliably despite message loss \cite{Computer-20:PoT,Computer-21:Bungie}.
Thus an unordered, lossy transport like UDP is sufficient for enacting BSPL protocols.

\emph{MAS Info (Configuration).} A protocol specifies a MAS abstractly via reference to roles. A \emph{concrete} MAS for a protocol is identified by a UUID and assigns roles to the agents that will play them.
MAS identifiers are essential since an agent may play a role in several MAS.
The properties of a (concrete) MAS and the mailboxes of the agents in the MAS are common knowledge to the agents in the MAS.
Kiko requires each agent to be configured with such knowledge;  Listing~\ref{lst:agent-config} gives such a configuration for agent \val{Bob}.

\begin{lstlisting}[caption={Bob's MAS Info Configuration.},label={lst:agent-config}]
self = "Bob"
systems = {
  "5feceb66": {
    "protocol": Purchase,
    "roles": {Buyer: self, Seller: "Sally"}}}
agents = {
  self: [("192.168.1.100", 1111)]
  "Sally": [("192.168.1.102", 1111), ("152.1.27.202", 1111)]}
\end{lstlisting}

In Listing~\ref{lst:agent-config}, \val{5feceb66} is an identifier for a MAS that enacts \pname{Purchase} with \val{Bob} and \val{Sally} as \rname{Buyer} and \rname{Seller}, respectively. \rname{Bob}'s and \rname{Sally}'s mailboxes are given as (IP, port) tuples. An agent may have several mailboxes for receiving messages; in Listing~\ref{lst:agent-config}, \rname{Sally} has two.
Our focus is not on how a MAS is constituted, but on programming abstractions that enable decentralized decision-making.
Listing~\ref{lst:agent-config} shows the kind of information needed to configure a MAS, and it could be constructed dynamically at runtime.

Formally, we model an agent using a tuple $\langle a, H_a, I_a, O_a \rangle$, where the components are the name of the agent, its history, input channel (its mailbox), and output channel respectively. Channels $I_a$ and $O_a$  are simply sets of message instances being sent and received, respectively, by agent $a$.  Definition~\ref{def:MAS} defines a MAS.

\begin{definition}[MAS]
\label{def:MAS}
A multiagent system $\mu$ is a tuple $\langle P, A \rangle$, where $P$ is a protocol, and $A$ is a map from roles of $P$ to agents.
\end{definition}

\emph{Decision Makers.} To write an agent, programmers supply the configuration and  write one or more decision makers.
A decision maker is invoked upon the occurrence of specified events.
When invoked, the adapter supplies it with prototypes of message instances that the agent is enabled to send given the agent's current history of message observations.
We refer to these prototypes as \emph{forms}, after documents with fields that need to be filled.
A form of a message schema has bindings for the parameters that are adorned $\inn$ in the schema, reflecting that its causal dependencies are satisfied, leaving only the parameters adorned \out to be bound.
The purpose of a decision maker is to flesh out some message instances from the forms by supplying bindings for their \out parameters; the adapter collects this set of completed instances as an emission \emph{attempt}.
The adapter verifies whether the attempt as a whole is consistent with the agent's history and if so, emits the instances in the attempt; else it rejects the attempt. 

Suppose \val{Bob}'s history is empty (it has observed no messages). Then the only form available to \val{Bob} is \val{Bob} -> \val{Sally}: \mname{RFQ}[\val{5feceb66}, (\paraname{ID}), (\paraname{item})], with unfilled parameters in parentheses.
Since protocol enactments occur within the context of a MAS, each form and any instance produced from it contains a MAS identifier (here, \val{5feceb66})---conceptually like the value for an implicit parameter \paraname{system} in every message.
Bob's programmer may have written a decision maker that fleshes out the above form into instances such as \val{Bob} -> \val{Sally}: \mname{RFQ}[\val{5feceb66}, \val{1}, \val{bat}] and \val{Bob} -> \val{Sally}: \mname{RFQ}[\val{5feceb66}, \val{2}, \val{ball}] based on some decision logic.
These instances are passed on to the adapter for emission. 
Listing~\ref{BSPL:buyer-init} shows a decision maker (in Python) called \val{start} that is invoked at system initialization, upon \val{InitEvent}. 
The argument \val{enabled} contains the available forms when \val{start} is invoked and the body of \val{start} contains code to send two instances of the form, one each for \val{bat} and \val{ball}.
The instruction to the adapter to emit the instances is implicit---after the decision maker returns, the adapter goes through all forms to see which ones have been fleshed out into instances and emits them (conditional to validation).

\begin{lstlisting}[caption={Bob's initial decision to send RFQs.},label={BSPL:buyer-init}]
@adapter.decision(event=InitEvent)
def start(enabled):
    for item in ["ball", "bat"]:
        ID = str(uuid.uuid4())
        for m in enabled.messages(RFQ):
            m.bind(ID=ID, item=item)
\end{lstlisting}

Consider another example. Suppose \val{Bob}'s history contains the above two RFQ instances and \val{Sally} -> \val{Bob}: \mname{Quote}[\val{5feceb66}, \val{1}, \val{bat}, \val{5}].  Then,  in addition to the \mname{RFQ} form specified above, the following forms would also be available to \val{Bob}: \val{Bob} -> \val{Sally}: \mname{Buy}[\val{5feceb66}, \val{1}, \val{bat}, \val{5}, (\paraname{done})] and \val{Bob} -> \val{Sally}: \mname{Reject}[\val{5feceb66}, \val{1}, \val{bat}, (\paraname{done})].  \val{Bob}'s programmer may have implemented a decision maker (as illustrated in Listing~\ref{lst:Bob-buy-reject}) that chooses from one of these two available forms based on how acceptable the \paraname{price} is, fleshes it out by binding \paraname{done}, and instructs the adapter to emit the resulting instance.

\begin{lstlisting}[caption={A simple Buy or Reject decision maker for Bob.},label={lst:Bob-buy-reject}]
@adapter.decision
def start(enabled):
    for m in enabled.messages(Buy):
        if(m["price"] < 20)
            m.bind(done="cool")
        else
            reject = next(enabled.messages(Reject, ID=m["ID"]))
            reject.bind(done="rejected")
\end{lstlisting}

We now give an example where a decision maker's emission attempt fails because it erroneously contains incompatible instances.
Specifically, Listing~\ref{lst:contradiction} is erroneous because Bob creates instances for both \mname{Buy} and \mname{Reject} in the same enactment.
This emission attempt fails because \mname{Buy} and \mname{Reject} are mutually exclusive according to Listing~\ref{BSPL:Purchase} (because both bind $\out$ \paraname{done}); neither will be emitted.

\begin{lstlisting}[caption={Decision maker attempting to send \mname{Buy} and \mname{Reject}.},label={lst:contradiction}]
@adapter.decision
def indecisive(enabled):
  buy = next(enabled.messages(Buy))
  reject = next(enabled.messages(Reject, system=buy.system, ID=buy["ID"]))
  buy.bind(done="accepted")
  reject.bind(done="rejected")
\end{lstlisting}

Listing~\ref{lst:contradiction}'s error brings out a remarkable aspect of Kiko.
Kiko enables decision makers (programmers) to choose sets of instances to emit.
Whereas each of the instances in the set (e.g., \mname{Buy}) would be individually consistent and compatible with the history when the decision maker was invoked and therefore could be emitted by the adapter, collectively, the set of instances chosen by the decision maker could be internally incompatible (\mname{Buy} and \mname{Reject}) and therefore fail emission by the adapter.
By rejecting incompatible emission sets, the adapter guarantees that an agent will not make noncompliant emissions.

An alternative would be to limit a decision maker to work on at most one form at a time.
Then, its emission by the adapter would be guaranteed.
Such a decision maker is a special case for Kiko. 

A specific triggering event may be specified for a decision maker (e.g., \val{InitEvent} in Listing~\ref{BSPL:buyer-init}). If such a triggering event is not specified (e.g., as in Listing~\ref{lst:Bob-buy-reject}), the adapter automatically invokes the decision maker whenever a communication event occurs.
Event-based invocation enables some optimizations:
First, the agent need not poll to wait for enough information to make a decision; not polling may be seen as an extension of the pub/sub pattern because a decision can depend on multiple pieces of information from multiple sources.
Second, because all constraints are relative to an enactment, and communication events contain keys identifying their enactment, the enactment can be directly looked up, thus avoiding linear scans or joins across an entire database for validation.

However, there are cases where an agent may want to emit messages outside of reacting to a message observation (whether sent or received). For example, if the agent needs to make business decisions only once per day, then waiting and making them all as a batch could be more efficient and accurate. To support a wider variety of behavioral patterns, Kiko uses an internal event queue on which the developer can signal custom events, and decision makers can be registered with custom filters to select which events should trigger them.

We now formalize the concepts introduced in the above section.

An \emph{association} binds values to some subset of the parameters of a message schema.

\begin{definition}[Association]
\label{def:association}
If $\schema{m}$ is a schema in protocol $P$, then $\mathcal{M}_\schema{m}$ is a relation with attributes $\fn{payload}(\schema{m}) = \langle \mu, \sender_\schema{m}, \receiver_\schema{m}, \vec{i}_\schema{m}, \vec{o}_\schema{m} \rangle$, and $\mathcal{M}$ is the union of all such relations.
The parameter name $\mu$ refers to a multiagent system.
A tuple $\association{m}$ is an \emph{association} of schema $\schema{m}$ if and only if it is a tuple of parameter bindings $\langle b_p | p \in \fn{payload}(\schema{m})] \rangle$ in $\mathcal{M}_\schema{m}$.
\end{definition}

We use $\association{m}[...]$ for projecting parameters to their bindings in the message instance; e.g., $\association{m}[\sender_\schema{m}]$ is the sender of $\association{m}$, and $\association{m}[\vec{k}_\schema{m}]$ is the projection of $\association{m}$'s key parameters.

A \emph{message instance} is an association where all parameters are bound.

\begin{definition}[Message Instance]
\label{def:instance}
An association $\association{m} \in \mathcal{M}_\schema{m}$ is a  \emph{message instance} and $\fn{instance}(\association{m})$ holds if and only if all of its parameters are bound: $p \in \fn{payload}(\association{m}), \association{m}[p] \neq \varnothing$.

$\mathcal{I} \subset \mathcal{M}$ is the set of all instances.
\end{definition}

A \emph{form} is an association where the \out parameters are unbound.

\begin{definition}[Form]
\label{def:form}
An association $\association{m} \in \mathcal{M}_\schema{m}$ is a \emph{form} (referring to a document with empty fields that need to be filled) if some \out parameter has a null value.
That is, $\forall p \in \fn{payload}(m)\setminus\vec{o}_\schema{m}: p \neq \varnothing$ and $\exists p \in \vec{o}_\schema{m}\colon \association{m}[p] = \varnothing$

$\mathcal{F} \subset \mathcal{M}$ is the set of all forms.
\end{definition}

We introduce the notion of context to capture enactments within a specific MAS.

\begin{definition}[Context]
\label{def:context}
The \emph{context} of an association is its MAS and its keys: $\association{m}[\mu, \vec{k}_m]$. 
\end{definition}

Associations \emph{share context} if their MAS and any of their keys have the same bindings. A form is enabled when all of its \inn parameter bindings match those from observed instances that share context (consistency), and its \out and \nil parameters do not conflict with any observed instances (compatibility), as given by Definitions~\ref{def:consistent}---\ref{def:enabled}.

\begin{definition}[Consistent]
\label{def:consistent}
Let $M,N \subseteq \mathcal{M}$ be sets of associations; then $N$ is \emph{consistent} with $M$ (and $\fn{consistent}(N,M)$ holds) if and only if the \inn bindings in $N$ are the same as bindings from associations that share context in $M$:

$\forall \association{m} \in M, \association{n} \in N\colon \association{m}[\mu,\vec{k}_m \cap \vec{k}_n] = \association{n}[\mu,\vec{k}_m \cap \vec{k}_n] \implies$\\ $\association{m}[\fn{payload}(m) \cap \vec{i}_n] = \association{n}[\fn{payload}(m) \cap \vec{i}_n]$.
\end{definition}

\begin{definition}[Out-Compatible]
\label{def:out-compatible}
Let $M,N \subseteq \mathcal{M}$ be sets of associations; then $N$ is \emph{out-compatible} with $M$ (and $\fn{compatible}_{\vec{o}}(N,M)$ holds) if and only if no \out bindings in $N$ are in payloads of associations that share context in $M$:

$\forall \association{m} \in M, \association{n} \in N\colon \association{m}[\mu,\vec{k}_m \cap \vec{k}_n] = \association{n}[\mu,\vec{k}_m \cap \vec{k}_n] \implies \fn{payload}(m) \cap \vec{o}_{\association{n}} = \varnothing$
\end{definition}

\begin{definition}[Nil-Compatible]
\label{def:compatible}
Let $M,N \subseteq \mathcal{M}$ be sets of associations; then $N$ is \emph{nil-compatible} with $M$ (and $\fn{compatible}_{\vec{n}}(N,M)$ holds) if and only if no \nil bindings in $N$ are in payloads of associations that share context in $M$:

$\forall \association{m} \in M, \association{n} \in N\colon \association{m}[\mu,\vec{k}_m \cap \vec{k}_n] = \association{n}[\mu,\vec{k}_m \cap \vec{k}_n] \implies \fn{payload}(m) \cap \vec{n}_{\association{n}} = \varnothing$
\end{definition}

\begin{definition}[Derived]
\label{def:derived}
Let $H_a$ be an agent history and $\association{m}$ be a form whose sender is $a$; then $\association{m}$ is \emph{derived from $H_a$} (and $\fn{derived}(\association{m}, H_a)$ holds) if and only if all of $\association{m}$'s \inn parameters are drawn from instances that share context in the history:
    
    $\forall p \in \vec{i}_{\schema{m}}, \exists \association{n} \in H_a\colon \association{n}[\mu,\vec{k}_m \cap \vec{k}_n] = \association{m}[\mu,\vec{k}_m \cap \vec{k}_n] \land p \in \vec{i}_n \land \association{m}[p] = \association{n}[p]$
\end{definition}

\begin{definition}[Enabled]
\label{def:enabled}
A message form $\association{m}$ is \emph{enabled} and $\fn{enabled}(\association{m},a,H_a)$ holds if and only if:
\begin{enumerate}
    \item $\association{m}$ is sent by $a$: $\association{m}[\sender_m] = a$
    \item $\fn{consistent}(\{\association{m}\}, H_a)$
    \item $\fn{compatible}_{\vec{o}}(\{\association{m}\}, H_a) \land \fn{compatible}_{\vec{n}}(\{\association{m}\}, H_a)$
    \item $\fn{derived}(\association{m}, H_a)$
\end{enumerate}

We also say that $\fn{enabled}(a,H_a) \subset \mathcal{F}$ is the set of message forms that $a$ is enabled to send.
\end{definition}

Definition~\ref{def:decision-maker} says a decision maker constructs only instances that preserve the bindings from message forms.

\begin{definition}[Decision Maker]
\label{def:decision-maker}
Let $Q$ be a set of message forms; a \emph{decision maker} is a function $d\colon \mathcal{P}(\mathcal{F}) \to \mathcal{P}(\mathcal{I})$ such that $\association{m}' \in d(Q) \implies \fn{instance}(\association{m}') \land \exists \partt{m} \in Q\colon \association{m}'[\sender_m, \receiver_m, \vec{i}_m] = \partt{m}[\sender_m, \receiver_m, \vec{i}_m]$.
\end{definition}

\subsection{Decision-Making Challenges and Solutions}

We highlight select decision making patterns supported by Kiko.

\subsubsection{Correlation}
An agent may simultaneously be involved in several enactments of a protocol. For example, \rname{Buyer} may be concurrently engaged with \rname{Seller} in several distinct enactments, each for some item at some price. The programming model should enable correlating communications by enactment.

Kiko supports correlation through the automatic derivation of correlated forms by the adapter (as described above).  The adapter computes forms based on all information available, potentially from the observation of multiple correlated instances.  Kiko also makes it convenient to find correlated forms where the decision logic requires it. For example, in Listing~\ref{lst:Bob-buy-reject}, correlated Reject forms are found by the \paraname{ID} of the Buy forms.

\subsubsection{Cross-Enactment Decisions}

Agents should be able to use information across enactments in their decision making. 

Kiko enables cross-enactment reasoning by providing forms from \emph{all} currently active \emph{contexts}, that is, enactments in all systems, to the decision makers together. Thus, the decision maker can select forms from multiple contexts and flesh them out for emission. For example, \val{Bob} could participate in multiple systems, all enacting \pname{Purchase}, to request quotes for the same item from multiple sellers. Then, \val{Bob} can send a \mname{Buy} for the \mname{Quote} with the lowest price (Listing~\ref{Bob:cheapest-buy}).  

\begin{lstlisting}[label={Bob:cheapest-buy}, caption={Selecting cheapest \mname{Buy} across multiple contexts.}]
@adapter.decision
def cheapest(enabled):
  buys = enabled.messages(Buy)
  cheapest = min(buys, key=lambda b: b["price"])
  cheapest.bind(done=True)
\end{lstlisting}

\subsubsection{Multiple Protocols}
An agent will often play roles in multiple unrelated protocols, using information from one to make decisions in another.  

Kiko enables implementing agents that play roles in multiple unrelated protocols.  For example, we specify \pname{Approval} in Listing~\ref{BSPL:Approval}. By enacting \pname{Approval} concurrently with \pname{Purchase}, \val{Bob} can seek \val{Alice}'s approval on any purchases.
To do so, \val{Bob} must map between the protocols inside its decision makers, which is supported by the enabled set containing forms from \emph{all} the protocols \val{Bob} is enacting.

\begin{lstlisting}[label={BSPL:Approval},caption={The \pname{Approval} protocol.}]
Approval {
  roles Requester, Approver
  parameters out aID key, out request, out approved
  
  Requester -> Approver: Ask[out aID key, out request]
  Approver -> Requester: Approve[in aID, in request, out approved]
}
\end{lstlisting}

Listing~\ref{lst:request-approval} shows \val{Bob}'s decision maker for constructing an \mname{Ask} (approval) for each \mname{Buy} as it becomes available as a form, copying \mname{Buy}'s payload into \paraname{request}. 

\begin{lstlisting}[caption={Requesting approval for a purchase across protocols.},label={lst:request-approval}]
@adapter.enabled(Buy)
def request_approval(buy):
    ask = next(adapter.enabled_messages.messages(Ask), None)
    return ask.bind(ID=str(uuid.uuid4()), request=buy.payload)
\end{lstlisting}

\subsubsection{Emission sets}

For additional flexibility, Kiko enables a decision maker to emit multiple instances atomically: if the instances are mutually compatible, then they are all emitted, else none are emitted.
Thus, e.g., if an emission set contained \mname{Buy} and \mname{Reject} instances for the same enactment, no instance in the set would be emitted.  Such atomicity of emission ensures correctness and gives full authority to the decision maker to choose its intended messages; multiple attempts can be made if needed.  Selecting some consistent subset of the emission set for emission, by contrast, would be arbitrary and could lead to unintended enactments.

Listing~\ref{lst:buyer-implementation} shows a decision maker, where \val{Bob} figures out the best combination of items it can buy (as computed by some optimization, whose details are not relevant for our purposes), sending \mname{Buy}s for all those items and \mname{Reject}s for the others. 

\begin{lstlisting}[caption={A decision maker that sends \mname{Buy} in some contexts and \mname{Reject}s in the others.},label={lst:buyer-implementation},numbers=left,xleftmargin=2em]
@adapter.decision
def select_gifts(enabled):
    best, rest = best_combo(enabled)
    for b in best:  # buy the best items
        b.bind(done=True)
    for r in rest:  # reject the rest
        r.bind(done=True)
\end{lstlisting}

Another variety of decision logic where emission sets are valuable is a combination of ``front-end'' and ``back-end'' reasoning.  For example, imagine \val{Sally} has a supplier with whom it engages via some protocol.  Suppose \val{Sally} wants to order an item from its supplier whenever it delivers an item to a buyer.  To accomplish this, it may have a decision maker which puts \mname{Deliver} (to the buyer) and \mname{Reorder} (from supplier) in the same emission set. 

\subsubsection{Reception-Order Freedom}

Requiring agents to receive messages in a particular order can only delay the reception of information, which in turn would limit the agent's ability to respond flexibly to events.  

Kiko takes advantage of the fact that BSPL doesn't rely on message ordering for correctness, and abstracts away message reception entirely from decision making.  An agent's adapter receives messages as they arrive and depending on the information in them, makes forms available to decision makers.  By doing so, Kiko enables agents to respond flexibly to events.

\begin{lstlisting}[caption={Rescind Quote.},label={BSPL:rescind-quote}]
  Seller -> Buyer: Rescind[in ID key, in item, in price, out rescinded]
  Buyer -> Seller: Buy[in ID key, ..., nil rescinded]
\end{lstlisting}

For example, Listing~\ref{BSPL:rescind-quote} extends \pname{Purchase} by allowing  \rname{Seller} to \mname{Rescind} a quote.
Because it depends on \paraname{price}, \mname{Rescind} must be sent after \mname{Quote}, but could reach \val{Bob} first. Because reception is not constrained except by integrity (inconsistent messages are rejected), \mname{Rescind} will be received, checked, and added to the history when it arrives.
As such, the matching \mname{Buy} will be disabled, and \val{Bob} need not waste any effort considering it (e.g., by requesting approval).

Note that by programming in terms of enabled forms, a decision maker such as the one in Listing~\ref{lst:Bob-buy-reject} that emits \mname{Buy}s need not change at all; the disabled \mname{Buy}s are simply not provided to the decision maker for consideration.

\subsubsection{Loose Coupling}

Clearly, protocols support the independent development of agents by capturing the constraints relevant to interoperation between them.  In general, if a protocol changes, then one would expect that the agents' decision making would have to change as well. Because Kiko is based on information though, it is not necessarily the case that protocol changes lead to changes in an agent's decision making, thus supporting loose coupling even better.

For example, suppose (as illustrated in Listing~\ref{BSPL:delivery}) \pname{Purchase} included a \mname{Deliver} message from \rname{Seller} that depended on \paraname{payment} provided by \mname{Buy}:

\begin{lstlisting}[caption={Delivery.},label={BSPL:delivery}]
  Buyer -> Seller: Buy[in ID key, in item, in price, out payment]
  Seller -> Buyer: Deliver[in ID key, in payment, out delivery]
\end{lstlisting}

Then, suppose \pname{Purchase} were extended so that \rname{Buyer} could pay indirectly via bank transfer (as illustrated in Listing~\ref{BSPL:bank-transfer}) .
Because the messages in Listing~\ref{BSPL:bank-transfer} do not change the messages emitted by \rname{Seller}, only how it receives the necessary information, \rname{Seller}'s decision logic need not be changed to support indirect payment. \rname{Seller}'s adapter will automatically derive the  \mname{Deliver} form when the indirect payment has been received, demonstrating loose coupling between the agents.

\begin{lstlisting}[caption={Bank Transfer.},label={BSPL:bank-transfer}]
  Buyer -> Seller: Accept[in ID key, in price, out acceptance]
  Buyer -> Bank: RequestTransfer[in ID key, in price, out txinfo]
  Bank -> Seller: Transfer[in ID key, in txinfo, out payment]
\end{lstlisting}

\subsubsection{Single Form Decision Makers}
The general decision making pattern of supporting the emission of sets of instances is highly flexible, but for cases in which an agent need emit only instance at a time, Kiko supports the convenient abstraction of \emph{single form} decision makers.
Such decision makers are functions invoked with a single message form; its return value is either a message instance for emission (binding its \out parameters), or a null value canceling the emission.  Listing~\ref{lst:quote-enablement} shows an example where an enabled form of \mname{Quote} is fleshed out.

\begin{lstlisting}[caption={Single Form Decision Maker for \mname{Quote}.},label={lst:quote-enablement}]
@adapter.enabled(Quote)
def send_quote(msg):
    msg["price"] = random.randint(20, 100)
    return msg
\end{lstlisting}

\subsection{Adapter Implementation}
\label{sec:impl}

\begin{figure}[htb!]
\centering
\resizebox{0.99\columnwidth}{!}{%
\definecolor{mpsNavy}{rgb}{0.01,0.01,0.5}
\definecolor{mpsGray}{rgb}{0.85,0.84,0.84}
\definecolor{mpsMyrtleGreen}{rgb}{0.19,0.47,0.45}
\definecolor{mpsRed}{rgb}{0.70,0.01,0.01}%
\begin{tikzpicture}
\usetikzlibrary{calc}
\usetikzlibrary{fit}
\usetikzlibrary{positioning}
\usetikzlibrary{matrix}
\tikzstyle{every text node part/.style}=[align=center]
\tikzset{>=latex}		

\tikzstyle{module}=[minimum width=100,inner sep=0,font=\small\sffamily,text=mpsNavy,fill=mpsGray,align=center,sharp corners,minimum height=15] 

\tikzstyle{wide}=[module,minimum width=110] 

\tikzstyle{emptybox}=[draw=none,fill=none,font=\small\sffamily,text=mpsNavy,minimum height=15]

\tikzstyle{framing}=[draw=mpsMyrtleGreen,thick,inner sep=2]
\tikzstyle{framing-highlight}=[framing,draw=mpsRed]

\tikzstyle{tier}=[draw=none,fill=none,font=\footnotesize\sffamily,text=red]%

\node [module] (receiver) {Receiver};
\node [framing,fit=(receiver)] (receiver-frame) {};

\node [module,right=1.9 of receiver] (emitter) {Emitter};
\node [framing,fit=(emitter)] (emitter-frame) {};

\coordinate (bottom-middle) at ($(emitter.east)!0.5!(receiver.west)$);

\coordinate (left-spot) at ($(receiver.west) + (0.01,0)$) {};
\coordinate (right-spot) at ($(emitter.east) + (-0.01,0)$) {};

\node [wide,fit=(left-spot)(right-spot),text height=1ex,above=1.1 of bottom-middle] (checker) {Checker};
\node [framing,fit=(checker)] (checker-frame) {};

\draw [mpsNavy,->] (receiver-frame.north) -- node[emptybox,right=0.05] {Receptions} (checker-frame.south-|receiver-frame.north);
\draw [mpsNavy,->] (checker-frame.south-|emitter-frame.north) -- node[emptybox,left=0.05] {Instances} (emitter-frame.north);


\node [module,above=3.5 of receiver.west,anchor=west,minimum height=25,minimum width=60] (local) {Local Store};
\node [framing,fit=(local)] (local-frame) {};

\draw [mpsNavy,<-] ($(local-frame.south)-(0.1,0)$) -- coordinate (ls-c) ($(checker-frame.north-|local-frame.south)-(0.1,0)$);
\draw [mpsNavy,->] ($(local-frame.south)+(0.1,0)$) -- coordinate (c-ls) ($(checker-frame.north-|local-frame.south)+(0.1,0)$);

\node[emptybox,left=0.1 of ls-c,align=center] (valid) {Valid\\Instances};
\draw ($(valid.east)-(0.4,0)$) -- (ls-c);
\node[emptybox,right=0.5 of c-ls] (history) {History};
\draw (history) -- (c-ls);

\node [module,right=1.25 of local.east,minimum height=25,minimum width=60] (mc) {Enablement};
\node [framing,fit=(mc)] (mc-frame) {};

\draw [mpsNavy,->] (local-frame.east) -- coordinate (ls-mc) (mc-frame.west);

\node [module,right=1.25 of mc.east,minimum height=25,minimum width=60] (dl) {Decision Makers};
\node [framing-highlight,fit=(dl)] (dl-frame) {};

\draw [mpsNavy,->] (mc-frame.east) -- coordinate (mc-dl) (dl-frame.west);
\draw [mpsNavy,->] (dl-frame.south) -- coordinate (dl-c) (checker-frame.north-|dl-frame.south);

\node[emptybox,above=0.45 of ls-mc] (enactments) {Enactments};
\draw (enactments) -- (ls-mc);
\node[emptybox,above=0.45 of mc-dl] (drafts) {Forms};
\draw (drafts) -- (mc-dl);
\node[emptybox,left=0.45 of dl-c] (attempts) {Attempts};
\draw (attempts) -- (dl-c);
\end{tikzpicture}
 
}
\caption{Adapter implementation.}
\label{fig:adapter-impl}
\end{figure}
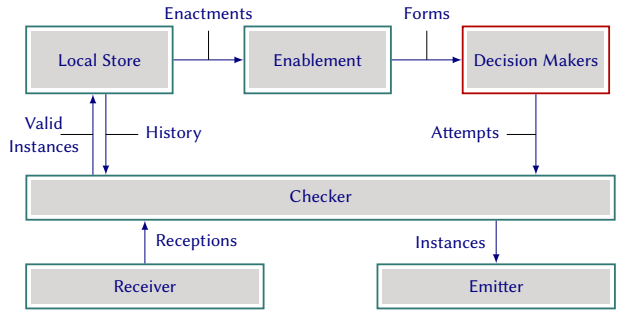

Figure~\ref{fig:adapter-impl} blows up the adapter from Figure~\ref{fig:kiko-agent} to highlight its internal components (highlighted in green).

The Emitter and Receiver interface with the communication service, putting messages on and receiving them from the wire, respectively. The Local Store records the agent's history of emissions and receptions.  The Checker validates (checking for satisfaction of causality and integrity constraints in the protocol specifications) any attempt (by a decision maker) to emit a set of messages (Definition~\ref{def:send-check}).  If an attempt is validated, then the instances in it are added to the Local Store and passed on the Emitter for emission; else, the attempt is discarded.  

\begin{definition}[Send-Check]
\label{def:send-check}
If $H_a \subseteq \mathcal{M}$ is a history for agent $a$, and $T \subseteq \mathcal{M}$ is a set of message instances, $\fn{check}_s(T, H_a)$ holds if and only if:
\begin{enumerate}
    \item $a$ is enabled to send every $\association{m}$ in $T$:
    
    $\forall \association{m} \in T, \fn{enabled}(\association{m}, a, H_a)$
    
    \item $T$ is out-, and nil-compatible with $T$:
    
    $\fn{compatible}_{\vec{o}}(T, T) \land \fn{compatible}_{\vec{n}}(T,T)$
\end{enumerate}
If $\fn{check}_s(T, H_a)$ holds, then $T$ is a valid set of emissions for $a$ and thus a valid extension of $H_a$.
\end{definition}

The Checker also validates received messages for integrity; if they pass, they are added to the Local Store, else they are discarded (Definition~\ref{def:receive-check}).

\begin{definition}[Receive-Check]
\label{def:receive-check}
If $H_a \subseteq \mathcal{M}$ is a history for agent $a$, and $\association{m} \in \mathcal{M}$ is a message instance, $\fn{check}_r(\association{m}, H_a)$ holds if and only if:
\begin{enumerate}
    \item \label{clause:check-receive} $\association{m}$ is receivable by $a$: $a = \association{m}[\receiver_m]$
    
    \item \label{clause:check-consistent} $\association{m}$ is consistent and out-compatible with the history:
    
    $\fn{consistent}(\{\association{m}\}, H_a) \land \fn{compatible}_{\vec{o}}(\{\association{m}\}, H_a)$
\end{enumerate}
If $\fn{check}_r(T, H_a)$ holds, it is valid for $a$ to receive every instance in $T$ and $T$ is a valid extension of $H_a$.
\end{definition}

The Local Store is used by  Enablement to compute the forms that the agent is enabled to send. Algorithm~\ref{algo:derive} describes how enabled forms are computed for each context. We use an incremental method, so that only those contexts that have new information are updated. First, on Line~\ref{line:derive-matching-contexts}, every context that shares key bindings with the observed instance $\association{o}$ is checked to see if it enables any instances of $m$. Lines~\ref{line:derive-check-outs} and~\ref{line:derive-check-nils} check that the \out and \nil parameters of the schema, respectively, are not already bound in the context. Line~\ref{line:derive-check-ins} copies the bindings of the \inn parameters from the context, Line~\ref{line:derive-system} copies the system ID, and Line~\ref{line:derive-add-form} adds the form to the result set for processing by decision makers.

\begin{algorithm}
\SetKwInOut{Input}{Input}
\Input{Message schema $m$, Message instance $\association{o}$}
$Q \gets \{\}$\;
\nl \label{line:derive-matching-contexts}\ForEach{$c \in matching\_contexts(\association{o})$}{
\nl \label{line:derive-check-outs} $o \gets \not\exists p\colon p \in \vec{o}_m \land p \in c.\fn{bindings}$\;
\nl \label{line:derive-check-nils} $n \gets \not\exists p\colon p \in \vec{n}_m \land p \in c.\fn{bindings}$\;
\nl \label{line:derive-check-ins} $i \gets \forall p\colon p \in \vec{i}_m \implies p \in c.\fn{bindings}$\;
    \If {$o \land i \land n$} {
\nl \label{line:derive-ins} $\association{m}[\vec{i}_m] \gets c.\fn{bindings}[\vec{i}_m]$\;
\nl \label{line:derive-system} $\association{m}[\mu] \gets \association{o}[\mu]$\;
\nl \label{line:derive-add-form} $Q \gets Q \cup \association{m}$\;
    }
    \Return{Q}\;
}
\caption{Derive instance of schema from observation.}
\label{algo:derive}
\end{algorithm}

\section{Operational Semantics}

Protocols are formalized in an online Appendix. Here, we formalize an agent and MAS computations via a transition semantics. 

\begin{figure}[htb]
\centering
\begin{mdframed}
\begin{subfigure}[c]{0.98\columnwidth}%
\begin{tabular}{lr@{$\mskip\thickmuskip$}c@{$\mskip\thickmuskip$}ll}
  Message Schema & $\schema{m}$ &$\in$& $S_P$ \\
  Message Instance & $\association{m}$ &$\in$& $\mathcal{M}$ \\
  History  & $H$ & $\in$ & $\mathcal{H} \subseteq \mathcal{M}$\\
  Input    & $I$     & $\subseteq$ & $\mathcal{M}$\\
  Output   & $O$     & $\subseteq$ & $\mathcal{M}$\\
  Agent    & $a$       & $\coloneq$  & $\langle H_a, I_a, O_a \rangle \in \mathcal{A}$ \\
  Check    & $\fn{check}_{r}$ & $\in$  & $\mathcal{M} \times \mathcal{H} \to \{\true, \false\}$ \\
           & $\fn{check}_{s}$ & $\in$  & $\mathcal{P}(\mathcal{M}) \times \mathcal{H} \to \{\true, \false\}$ \\
  Enabled  & $\fn{enabled}$ & $\in$ & $\mathcal{A} \times \mathcal{H} \to \mathcal{P}(\mathcal{F})$ \\
  Decision maker  & $d$ & $\in$ & $\mathcal{P}(\mathcal{F}) \to \mathcal{P}(\mathcal{M})$\\
  Consistent & $\fn{consistent}$ & $\in$ & $\mathcal{P}(\mathcal{M}) \to \{\true, \false\}$
\end{tabular}
\end{subfigure}

\hspace*{1em}
\begin{subfigure}[c]{0.98\columnwidth}%
    \begin{mathpar}
    \inferrule*[Left={Recv}]
    {\association{m} \in I_a\\
    \association{m} \not\in H_a\\
    \fn{check}_r(\association{m}, H_a)}
    {a\langle H_a, I_a, O_a \rangle
    \longrightarrow
    a \langle H_a \cup \{\association{m}\}, I_a, O_a \rangle}
    \end{mathpar}
    \begin{mathpar}
    \inferrule*[Left={Tx}]
    {\association{m} \in O_x\\
    \association{m}[\receiver]=y}
    {I_y \longrightarrow I_y \cup \{\association{m}\}} 
    \end{mathpar}
    \begin{mathpar}
    \inferrule*[Left={Decide}]
    {Q:=\fn{enabled}(a,H_a)\\
     T := d(Q)\\
    \fn{check}_s(T, H_a)}
    {a\langle H_a, I_a, O_a \rangle
    \longrightarrow
    a\langle H_a \cup T, I_a, O_a \cup T \rangle}
    \end{mathpar}
\end{subfigure}
\end{mdframed}
\caption{Notation and core semantics.}
\label{tab:-semantics}
\end{figure}

\begin{figure}[htb]
\centering
\begin{mdframed}
\hspace*{2em}
\begin{subfigure}[c]{0.98\columnwidth}
    \begin{mathpar}\mprset{sep=0.7em}
    \inferrule*[Left={Decide$_2$}]
    {Q := \fn{enabled}(a,H_a)\\
    T := d(Q)\\\\
    \fn{compatible}_{\vec{o}}(T,T)\\
    \fn{compatible}_{\vec{n}}(T,T)}
    {a\langle H_a, I_a, O_a \rangle
    \longrightarrow
    a\langle H_a \cup T, I_a, O_a \cup T \rangle}
    \end{mathpar}
\end{subfigure}
\end{mdframed}
\caption{Optimized decision that checks for internal consistency instead of full validity.}
\label{tab:alt-decide}
\end{figure}

Figure~\ref{tab:-semantics} gives the transition semantics.

The \fsc{Recv} rule specifies how messages are received.
For agent $a$ to receive a message instance $\association{m}$ there are three conditions:
\begin{enumerate*}
\item $\association{m}$ must be in the agent's input channel $I_a$,
\item \association{m} must not already be in the agent's history $H_a$, and
\item $\association{m}$ must be a valid extension of $H_a$.
\end{enumerate*}
If these three conditions are met, then \association{m} is added to $H_a$.

The \fsc{Tx} rule models message delivery by copying messages from an output channel to the appropriate input channel; unreliability is modeled by not exercising the rule.

Finally, \fsc{Decide} specifies how messages are instantiated for emission:
First, a set $Q$ of message forms is computed based on the agent's history.
Next, a set of instances are derived from the message forms by applying a decision maker $d$ to the enabled form set.
If this set of instances is valid, then it is added to both the agent's history and output channel.
Otherwise, the rule cannot be applied and no messages are sent.

No rules are required for cases where the messages fail a validity check; there is simply no transition in those cases. A transition for a MAS is simply a transition for one of its agents.

Figure~\ref{tab:alt-decide} shows an alternative version of the \fsc{Decide} rule, $\fsc{Decide}_2$.  Because transitions are atomic, the forms will not be disabled before the transition completes, so they do not need to be rechecked for validity; checking internal compatibility is sufficient (e.g., not selecting both an \mname{Accept} and \mname{Reject} in the same enactment).
Checking only internal compatibility of a small set of emissions should be faster than a full send-check, which requires both internal compatibility \emph{and} that the instance is consistent and compatible with the rest of the agent's history.

Our goal is to show that a MAS developed using our operational semantics to implement a protocol will be both correct (that is, reach only valid states) and complete (it is possible to implement a system that can reach any valid state).
As such, we formalize the state of a MAS, which states are reachable according to the operational semantics, and which states \emph{match} a protocol enactment.

\begin{definition}[MAS State]
\label{def:mas-state}
The \emph{state} of a MAS $\mu$ is the set of its agent histories: $\{H_a | a \in A_\mu \}$
\end{definition}

\begin{definition}[Reachable State]
Given MAS $\mu$ and transition semantics $\mathcal{T}$, state $s$ of MAS $\mu$ is \emph{reachable} and an element of $\mathcal{S}_{\mu,\mathcal{T}}$ if and only if there is a sequence of transitions $t_i \in \mathbb{N} \to \mathcal{T}$ that results in state $s$.
\end{definition}

$\mathcal{E}_P$ (formally defined in the appendix) is the set of reachable enactments of protocol $P$, where a reachable enactment $E \in \mathcal{E}_P$ is a set of role histories each constructed by a sequence of viable events according to $P$'s specification.

\begin{definition}[Matching State]
If $\mu$ is a MAS implementing protocol $P$, then state $s$ of $\mu$ \emph{matches} $E \in \mathcal{E}_P$, written $s \equiv E$, if and only if, for every agent history $H_a$ in $s$ and instance $\association{m} \in H_a$:
\begin{enumerate}
    \item if $a$ plays $\sender_m$ in $\mu$ then $m$ is sent in the corresponding role history $H_{\sender_m} \in E$ (that is, $a = \association{m}[\sender_m] \implies \langle \fn{sent}, m \rangle \in H_{\receiver_m}$)
    \item if $a$ plays the receiver of $m$ in $\mu$ then $m$ is received in the corresponding role history $H_{\receiver_m} \in E$ (that is, $a = \association{m}[\receiver_m] \implies \langle \fn{received}, m \rangle \in H_{\receiver_m}$)
\end{enumerate}
\end{definition}

Simulation is the idea that transitions in the MAS should match the reachable enactments in its protocol; each transition may be equivalent to a set of \emph{multiple} viable extensions because the \fsc{Decide} rule can produce a \emph{set} of message instances, where viable extensions cover only one instance at a time.

\begin{definition}[Simulation]
If $\mu$ is a MAS implementing protocol $P$, then state $s \in \mathcal{S}_{\mu}$ \emph{simulates} $E \in \mathcal{E}_P$, written $s \sim E$, if and only if, for every agent history $H_a$ in $s$ and instance $\association{m} \in H_a$:
\begin{enumerate}
    \item $s$ matches $E$
    \item for every transition $t$, the state $s'\colon s \overset{t}{\to} s'$ matches some enactment $E'$ reachable from $E$ in a finite number of viable extensions.
\end{enumerate}
\end{definition}

Theorem~\ref{thm:-simulation} gives the \emph{correctness} of our operational semantics by showing that compliant MAS can only reach states that match reachable enactments of a protocol.
Even though the states reached by the MAS will depend on the decision makers, they can only select subsets of the enabled forms, and therefore cannot reach an invalid state (that is, one that does not match an enactment that is reachable under the protocol semantics).

\begin{theorem}
\label{thm:-simulation}
Given a MAS $\mu$ implementing protocol $P$, every reachable state $s \in \mathcal{S}_{\mu}$ simulates some enactment $E \in \mathcal{E}_P$.
\end{theorem}

Theorem~\ref{thm:internal-checking-optimization} shows that the conditions for \fsc{Decide} are redundant, given that the forms are drawn from $\fn{enabled}(a,H_a)$ and decision makers preserve their bindings (and thus consistency and compatibility with history); all that needs to be checked for the selected emissions $T$ is that they are compatible with each other.

\begin{theorem}
\label{thm:internal-checking-optimization}
$\fsc{Decide}_2$ is equivalent to \fsc{Decide}.
\end{theorem}

Theorem~\ref{thm:completeness} shows \emph{completeness} for our operational semantics: the operational semantics do not restrict a MAS from simulating any reachable enactment of the protocol.
Or, given a reachable enactment of a protocol, it is possible to construct decision makers for the agents that would reach that enactment.
This is not to say that every implementation is complete; proving completeness for a given implementation would require formalizing its decision makers as transition rules.

\begin{theorem}
\label{thm:completeness}
Given a MAS $\mu$ implementing protocol $P$, there is some set of decision makers $D$ that can simulate any reachable enactment in $\mathcal{E}_P$, assuming that all sent message instances are received.
\end{theorem}

The proofs of these theorems are in the appendix.

\section{Discussion}
\label{sec:discussion}

Kiko bridges business logic and communications: an agent provides business decisions and the underlying adapter applies the protocol semantics to determine which messages are viable. The underlying causal information semantics captures the information flow and avoids having to generate guards \cite{Icde96}. 
An agent makes and communicates a set of decisions (as reflected in the forms provided by the adapter) based on some evaluation of the state of the world.  
The decision making is conceptualized declaratively and suits rule-based programming languages such as Jason.  

An interesting direction is to extend Kiko's notion of forms to support norms-based decision making.  For example, the discharge of a commitment by an agent could be made available as a form to be picked and instantiated by the agent. \citet{Baldoni+21:accountability} present a model for accountability that is implemented in JaCaMo via obligations and relates to both (the giving of) accounts and recovery strategies when things go wrong.  Kiko's adapter could incorporate standard protocols for demanding accounts from other agents when norm violations occur and incorporate them into further decision making, e.g., to decide from which agent to buy items.  

Variants of programming models based on information protocols have been proposed in recent years.  The idea of enabled message forms was first introduced in Stellar \cite{Gunay+Chopra-18:Stellar}; however, Stellar lacked support for emission sets and relied on the abstraction of message handlers as opposed to decision makers.  Thus, Bob's implementation in Stellar would be a set of message handlers, one for each type of message it could observe. Within a message handler, one could retrieve a form and instantiate it. Message handling-based abstractions are lower level compared to Kiko's decision makers, which are information-based.  To see this, suppose an agent needed information from two instances, say $i_1$ and $i_2$, which it may receive in any order, to be able to send a third instance $i_3$ (e.g., a shipper may need the address from the buyer and the item from the seller to be able to deliver).  Then, in the message handling approach of Stellar, one would write separate message handlers for $i_1$ and $i_2$ and in each one check whether the form for $i_3$ is available. By contrast, in Kiko, one would simply write a single decision maker that completes the form for $i_3$.  The Mandrake \cite{JAAMAS-22:Mandrake} and PoT \cite{Computer-20:PoT} programming models share Stellar's limitations; however, they both also address application-level fault tolerance, a theme that is a direction for Kiko.

Like Stellar (and Mandrake and PoT), Kiko enables building applications directly over an unordered, unreliable communication service such as UDP for message transport. Kiko is therefore compatible with the influential end-to-end argument \cite{Saltzer+84}, which advocates building applications over simple communication services, both for reasons of enabling application-level flexibility and performance. By contrast, message ordering-based protocol approaches would be incompatible with the end-to-end argument. Establishing the performance of Kiko-based agents and MAS compared to traditional application architectures that rely on complex communication services and middleware is a crucial direction.  Preliminary evidence from Mandrake and PoT indicates high performance.

Kiko's features such as support for correlation, cross-enactment reasoning, and multiple protocols are not readily supported in programming models for message ordering-based protocol approaches.  This is because all of the above features have to do with querying information, which is inadequately represented in ordering-based protocols. Emission sets are unique to Kiko and are a powerful feature that enables emitting a set of message instances (possibly from different protocols and to different agents) atomically.  

In our semantics, decision makers execute atomically with respect to the history, which simplifies checking the internal compatibility of the emission set before emitting all its instances.  However, an alternative semantics is possible where decision makers execute concurrently from the same history.  Concurrent execution would enable taking advantage of multicore and cloud architectures.  Implementation-wise, decision makers could be spawned off as actors \cite{Hewitt+73,Agha86}. The tradeoff is that the emission sets produced by concurrent decision makers may be in conflict with each other (e.g., one set contains \mname{Buy} whereas another contains \mname{Reject} for the same enactment) and therefore an internal compatibility check would no longer suffice.  Each emission set would have to be checked for validity against the history, which could be more expensive.

IoT-based paradigms such as edge and fog computing and the industry paradigm of realizing applications via microservices are conceptually decentralized.  In the case of microservices especially, decentralization is driven by the scalability afforded by the containerization of application components.  Current microservices development approaches tend to avoid distributed database transactions in favor of loose coupling \cite{Laigner+21:microservices}. However, this raises the question: On what basis should microservices coordinate their computations? Information protocols could be thought of as a model for \emph{business transactions}.  Therefore, approaches like Kiko, suitably adapted to microservices, can help.

SARL \cite{Galland+20:SARL}, an agent programming language, supports communication using events in spaces that are akin to environments \cite{Weyns+07:environment}. SARL would benefit from a protocol-based programming model. Kiko would benefit from a more general treatment of events. Currently, in Kiko, messages model events. However, some domain events don't map to messages. For example, while a \mname{Quote} may reasonably be modeled as a message, \mname{Shipment} may actually correspond to a package traveling in the back of a truck. Receiving a shipment, therefore, requires sensing the arrival of the package.  Extending Kiko's adapter to incorporate observation of events from the environment would be valuable. 

\emph{Supplementary Material.} The appendix and the Kiko software are available at: \url{https://gitlab.com/masr/bspl/-/tree/kiko}.

\clearpage

\section*{Acknowledgments} 

We thank the anonymous reviewers for their comments. We thank the EPSRC (grant EP/N027965/1) and the US National Science Foundation (grant IIS-1908374) for partial support.

\DeclareRobustCommand{\nUmErAL}[1]{#1}\DeclareRobustCommand{\nAmE}[3]{#3}


\end{document}